\documentstyle[12pt]{article}
\begin{document}

\title {THE ACTIVE GRAVITATIONAL MASS OF A HEAT CONDUCTING SPHERE
OUT OF HYDROSTATIC EQUILIBRIUM}
\author{L. Herrera\thanks{Postal address: Apartado 80793, Caracas 1080A,
Venezuela;
E-mail address: laherrera@telcel.net.ve}\,
and
A. Di Prisco$^*$\\
Escuela de F\'\i sica \\
Facultad de Ciencias\\
Universidad Central de Venezuela\\
Caracas, Venezuela.\\
}
\date{}
\maketitle

\begin{abstract}
We obtain an expression for the active gravitational mass
of a relativistic heat conducting fluid, just after its
departure from hydrostatic equilibrium, on a time scale of
the order of relaxation time.

It is shown that an increase of a characteristic parameter
leads to larger (smaller) values of active gravitational mass of
collapsing (expanding) spheres, enhacing thereby the instability
of the system.
\end{abstract}

\section{Introduction}
In a recent series of works \cite{Heetal97,HeMa971,HeMa972} the behaviour
of dissipative systems at
the very moment when they depart from hydrostatic equilibrium, has been studied.

\noindent
It appears that a parameter formed by a specific combination of thermal
relaxation
time, temperature, proper energy density and pressure, may critically
affect the
evolution of the object.

\noindent
More specifically, it has been shown that in the equation of motion of any
fluid element,
the inertial mass term is multiplied by a factor, vanishing for a given
value of that parameter
(critical point) and changing of sign beyond that value.

\noindent
Although the above mentioned parameter is constrained by causality
requirements, it appears that in some
cases these requirements do not prevent the system from reaching the
critical point
\cite{HeMa972}. Furthermore it might not be reasonable to apply, close to
the critical point,
restrictions obtained from a linear perturbative scheme (as is the case for
causality
conditions) \cite{HeMa971}.

\noindent
In order to delve more deeply into the physical nature of the critical
point, we shall
obtain here an expression for the active gravitational mass (Tolman mass
\cite{To})
which explicitly contains the parameter mentioned above.

\noindent
It will be seen that this expression yields larger (smaller) values for the
active
gravitational mass of the inner core of a collapsing (expanding) sphere, as
we approach
the critical point, this tendency persists beyond the critical point as the
systems moves
away from it.

\noindent
This result provides some hints about the way in which the evolution of the
system is
affected by the aforesaid parameter.

\noindent
The paper is organized as follows.

\noindent
In the next section the field equations, the conventions and other useful
formulae are introduced.
In section 3 we briefly present the equation for the heat conduction. The
departure from hydrostatic
equilibrium is considered in section 4. In section 5 we derive an
expression for the Tolman mass
and evaluate it at the very moment the system departs from hydrostatic
equilibrium . Finally a
discussion of this expression is presented in the last section.

\section{Field Equations and Conventions.}

\noindent
We consider spherically symmetric distributions of collapsing
fluid, which for sake of completeness we assume to be anisotropic,
undergoing dissipation in the form of heat flow, bounded by a
spherical surface $\Sigma$.

\noindent
The line element is given in Schwarzschild-like coordinates by

\begin{equation}
ds^2=e^{\nu} dt^2 - e^{\lambda} dr^2 -
r^2 \left( d\theta^2 + sin^2\theta d\phi^2 \right)
\label{metric}
\end{equation}

\noindent
where $\nu(t,r)$ and $\lambda(t,r)$ are functions of their arguments. We
number the coordinates: $x^0=t; \, x^1=r; \, x^2=\theta; \, x^3=\phi$.

\noindent
The metric (\ref{metric}) has to satisfy Einstein field equations

\begin{equation}
G^\nu_\mu=-8\pi T^\nu_\mu
\label{Efeq}
\end{equation}

\noindent
which in our case read \cite{Bo}:

\begin{equation}
-8\pi T^0_0=-\frac{1}{r^2}+e^{-\lambda}
\left(\frac{1}{r^2}-\frac{\lambda'}{r} \right)
\label{feq00}
\end{equation}

\begin{equation}
-8\pi T^1_1=-\frac{1}{r^2}+e^{-\lambda}
\left(\frac{1}{r^2}+\frac{\nu'}{r}\right)
\label{feq11}
\end{equation}

\begin{eqnarray}
-8\pi T^2_2  =  -  8\pi T^3_3 = & - &\frac{e^{-\nu}}{4}\left(2\ddot\lambda+
\dot\lambda(\dot\lambda-\dot\nu)\right) \nonumber \\
& + & \frac{e^{-\lambda}}{4}
\left(2\nu''+\nu'^2 -
\lambda'\nu' + 2\frac{\nu' - \lambda'}{r}\right)
\label{feq2233}
\end{eqnarray}

\begin{equation}
-8\pi T_{01}=-\frac{\dot\lambda}{r}
\label{feq01}
\end{equation}

\noindent
where dots and primes stand for partial differentiation with respect
to t and r
respectively.

\noindent
In order to give physical significance to the $T^{\mu}_{\nu}$ components
we apply the Bondi approach \cite{Bo}.

\noindent
Thus, following Bondi, let us introduce purely locally Minkowski
coordinates ($\tau, x, y, z$)

$$d\tau=e^{\nu/2}dt\,\qquad\,dx=e^{\lambda/2}dr\,\qquad\,
dy=rd\theta\,\qquad\, dz=rsin\theta d\phi$$

\noindent
Then, denoting the Minkowski components of the energy tensor by a bar,
we have

$$\bar T^0_0=T^0_0\,\qquad\,
\bar T^1_1=T^1_1\,\qquad\,\bar T^2_2=T^2_2\,\qquad\,
\bar T^3_3=T^3_3\,\qquad\,\bar T_{01}=e^{-(\nu+\lambda)/2}T_{01}$$

\noindent
Next, we suppose that when viewed by an observer moving relative to these
coordinates with proper velocity $\omega$ in the radial direction, the physical
content  of space consists of an anisotropic fluid of energy density $\rho$,
radial pressure $P_r$, tangential pressure $P_\bot$ and radial heat flux
$\hat q$. Thus, when viewed by this moving observer the covariant tensor in
Minkowski coordinates is

\[ \left(\begin{array}{cccc}
\rho    &  -\hat q  &   0     &   0    \\
-\hat q &  P_r      &   0     &   0    \\
0       &   0       & P_\bot  &   0    \\
0       &   0       &   0     &   P_\bot
\end{array} \right) \]

\noindent
Then a Lorentz transformation readily shows that

\begin{equation}
T^0_0=\bar T^0_0= \frac{\rho + P_r \omega^2 }{1 - \omega^2} +
\frac{2 Q \omega e^{\lambda/2}}{(1 - \omega^2)^{1/2}}
\label{T00}
\end{equation}

\begin{equation}
T^1_1=\bar T^1_1=-\frac{ P_r + \rho \omega^2}{1 - \omega^2} -
\frac{2 Q \omega e^{\lambda/2}}{(1 - \omega^2)^{1/2}}
\label{T11}
\end{equation}

\begin{equation}
T^2_2=T^3_3=\bar T^2_2=\bar T^3_3=-P_\bot
\label{T2233}
\end{equation}

\begin{equation}
T_{01}=e^{(\nu + \lambda)/2} \bar T_{01}=
-\frac{(\rho + P_r) \omega e^{(\nu + \lambda)/2}}{1 - \omega^2} -
\frac{Q e^{\nu/2} e^{\lambda}}{(1 - \omega^2)^{1/2}} (1 + \omega^2)
\label{T01}
\end{equation}

\noindent
with

\begin{equation}
Q \equiv \frac{\hat q e^{-\lambda/2}}{(1 - \omega^2)^{1/2}}
\label{defq}
\end{equation}

\noindent
Note that the coordinate velocity in the ($t,r,\theta,\phi$) system, $dr/dt$,
is related to $\omega$ by

\begin{equation}
\omega=\frac{dr}{dt}\,e^{(\lambda-\nu)/2}
\label{omega}
\end{equation}

\noindent
At the outside of the fluid distribution, the spacetime is that of Vaidya,
given by

\begin{equation}
ds^2= \left(1-\frac{2M(u)}{{\cal R}}\right) du^2 + 2dud{\cal R} -
{\cal R}^2 \left(d\theta^2 + sin^2\theta d\phi^2 \right)
\label{Vaidya}
\end{equation}

\noindent
where $u$ is a time-like coordinate such that $u=constant$ is
(asymptotically) a
null cone open to the future and ${\cal R}$ is a null coordinate ($g_{\cal
RR}=0$). It should
be remarked, however, that strictly speaking, the radiation can be considered
in radial free streaming only at radial infinity.

\noindent
The two coordinate systems ($t,r,\theta,\phi$) and ($u,{\cal
R},\theta,\phi$) are
related at the boundary surface and outside it by

\begin{equation}
u=t-r-2M\,ln \left(\frac{r}{2M}-1\right)
\label{u}
\end{equation}

\begin{equation}
{\cal R}=r
\label{R}
\end{equation}

\noindent
In order to match smoothly the two metrics above on the boundary surface
$r=r_\Sigma(t)$, we have to require the continuity of the first fundamental
form across that surface. As result of this matching we obtain

\begin{equation}
\left[P_r\right]_\Sigma=\left[Q\,e^{\lambda/2}\left(1-\omega^2\right)^
{1/2}\right]_\Sigma = \left[\hat q\right]_\Sigma
\label{PQ}
\end{equation}

\noindent
expressing the discontinuity of the radial pressure in the presence
of heat flow, which is a well known result \cite{Sa}.

\noindent
Next, it will be useful to calculate the radial component of the
conservation law

\begin{equation}
T^\mu_{\nu;\mu}=0
\label{dTmn}
\end{equation}

\noindent
After tedious but simple calculations we get

\begin{equation}
\left(-8\pi T^1_1\right)'=\frac{16\pi}{r} \left(T^1_1-T^2_2\right)
+ 4\pi \nu' \left(T^1_1-T^0_0\right) +
\frac{e^{-\nu}}{r} \left(\ddot\lambda + \frac{\dot\lambda^2}{2}
- \frac{\dot\lambda \dot\nu}{2}\right)
\label{T1p}
\end{equation}

\noindent
which in the static case becomes

\begin{equation}
P'_r=-\frac{\nu'}{2}\left(\rho+P_r\right)+
\frac{2\left(P_\bot-P_r\right)}{r}
\label{Prp}
\end{equation}

\noindent
representing the generalization of the Tolman-Oppenheimer-Volkof equation
for anisotropic fluids \cite{BoLi}.

\section{Heat Conduction Equation.}

\noindent
In the study of star interiors
it is usually assumed that the energy flux of radiation (and
thermal conduction) is proportional to the gradient of temperature
(Maxwell-Fourier law or Eckart-Landau in general relativity).

\noindent
However it is well known that the Maxwell-Fourier law for the radiation
flux leads to a parabolic equation (diffusion equation) which predicts
propagation of perturbation with infinite speed (see \cite{Jo}--\cite{Maa} and
references therein). This simple fact is at the origin of the pathologies
\cite{HiLi} found in the approaches of Eckart \cite{Ec} and Landau \cite{La}
for relativistic dissipative processes.

\noindent
To overcome such difficulties, different relativistic
theories with non-vanishing relaxation times have been proposed
in the past \cite{Is}--\cite{Car}. The important point is that all these
theories provide a heat transport equation which is not of
Maxwell-Fourier type but of Cattaneo type \cite{Cat}, leading thereby to a
hyperbolic equation for the propagation of thermal perturbation.

\noindent
Accordingly we shall describe the heat transport by means of a
relativistic Israel-Stewart equation \cite{Maa}. Although a complete
treatment of dissipative
processes requires the inclusion of viscous stresses as well as the
coupling between these
and the heat flow, we shall assume here for simplicity vanishing viscosity.
Thus we have

\begin{equation}
\tau \frac{Dq^\alpha}{Ds} + q^\alpha =
\kappa P^{\alpha \beta} \left(T_{,\beta} - T a_\beta\right) -
\tau u^\alpha q_\beta a^\beta-
\frac{1}{2} \kappa T^2
\left(\frac{\tau}{\kappa T^2} u^\beta\right)_{;\beta} q^\alpha
\label{Catrel}
\end{equation}

\noindent
with

\begin{equation}
u^\mu=\left(\frac{e^{-\nu/2}}{\left(1-\omega^2\right)^{1/2}},\,
\frac{\omega\, e^{-\lambda/2}}{\left(1-\omega^2\right)^{1/2}},\,0,\,0\right)
\label{umu}
\end{equation}

\begin{equation}
q^\mu=Q\,\left(\omega\,e^{(\lambda-\nu)/2},\,1,\,0,\,0\right)
\label{qmu}
\end{equation}

\noindent
where $\kappa$, $\tau$, $T$, $q^\beta$ and $a^\beta$ denote thermal
conductivity,
thermal relaxation time, temperature, the heat flow vector and the
components of the four
acceleration, respectively. Also, $P^{\alpha \beta}$ is the projector
onto the hypersurface orthogonal to the four velocity $u^\alpha$.

\section{Thermal Conduction and Departure from Hydrostatic Equilibrium.}

\noindent
Let us now consider a spherically symmetric fluid distribution which
initially may be in either hydrostatic and thermal equilibrium (i.e.
$\omega = Q = 0$), or slowly evolving and dissipating energy through
a radial heat flow vector.

\noindent
Before proceeding further with the treatment of our problem, let us
clearly specify the meaning of ``slowly evolving''. That means that
our sphere changes on a time scale which is very large as compared to
the typical time in which it reacts on a slight perturbation of
hydrostatic equilibrium. This typical time is called hydrostatic
time scale. Thus a slowly evolving system is always in hydrostatic
equilibrium (very close to), and its evolution may be regarded as
a sequence of static models linked by (\ref{feq01}).
This assumption is very sensible, since
the hydrostatic time scale is usually very small.

\noindent
In fact, it is of the order of $27$ minutes for the sun, $4.5$ seconds
for a white dwarf and $10^{-4}$ seconds for a neutron star of one
solar mass and $10$ Km radius \cite{Ki}.

\noindent
In terms of $\omega$ and metric functions, slow evolution means
that the radial velocity $\omega$ measured by the Minkowski observer,
as well as time derivatives are so small that their products and
second order time derivatives may be neglected (an invariant
characterization of slow evolution may be found in \cite{HeSa95}).

\noindent
Thus \cite{HeDP97}

\begin{equation}
\ddot\nu\approx\ddot\lambda\approx\dot\lambda \dot\nu\approx
\dot\lambda^2\approx\dot\nu^2\approx
\omega^2\approx\dot\omega=0
\label{neg}
\end{equation}

\noindent
As it follows from (\ref{feq01}) and (\ref{T01}), $Q$ is of the
order $O(\omega)$.
Therefore in the slowly evolving regime, relaxation terms may be neglected
and (\ref{Catrel}) becomes the usual Landau-Eckart transport equation
\cite{HeDP97}.

\noindent
Then, using (\ref{neg}) and (\ref{T1p}) we obtain (\ref{Prp}),
which as mentioned before is the equation of hydrostatic equilibrium
for an anisotropic fluid. This is in agreement with what was mentioned
above, in the sense that a slowly evolving system is in hydrostatic
equilibrium.

\noindent
Let us now return to our problem. Before perturbation, the two
possible initial states of our system are characterized by:

\begin{enumerate}
\item Static
\begin{equation}
\dot \omega = \dot Q = \omega = Q = 0
\label{eqdt}
\end{equation}
\item Slowly evolving
\begin{equation}
\dot \omega = \dot Q = 0
\label{evlen}
\end{equation}
\begin{equation}
Q \approx O(\omega) \not = 0 \; \qquad (small)
\label{Qorom}
\end{equation}
\end{enumerate}

\noindent
where the meaning of ``small'' is given by (\ref{neg}).

\noindent
Let us now assume that our system is submitted to perturbations
which force it to depart from hydrostatic equilibrium but keeping the
spherical symmetry.

\noindent
We shall study the perturbed system on a time scale which is
small as compared to the thermal adjustment time.

\noindent
Then, immediately after perturbation (``immediately'' understood
in the sense above), we have for the first initial condition
(static)

\begin{equation}
\omega = Q = 0
\label{omyQ0}
\end{equation}

\begin{equation}
\dot\omega \approx \dot Q \not = 0 \; \qquad (small)
\label{chiq}
\end{equation}

\noindent
whereas for the second initial condition (slowly evolving)

\begin{equation}
Q \approx O(\omega) \not = 0 \; \qquad (small)
\label{Qseg}
\end{equation}

\begin{equation}
\dot Q \approx \dot\omega \not = 0 \; \qquad (small)
\label{pomQ2}
\end{equation}

\noindent
As it was shown in \cite{Heetal97}, it follows from (\ref{T1p}) and
(\ref{T00})--(\ref{T01})
that after perturbation, we have for both initial conditions (see
\cite{Heetal97} for details)

\begin{equation}
- e^{(\nu-\lambda)/2} R = \left(\rho+P_r\right) \dot\omega +
\dot Q e^{\lambda/2}
\label{pfR}
\end{equation}

\noindent
where $R$ denotes the left-hand side of the TOV equation, i.e.

\begin{eqnarray}
R & \equiv &  \frac{dP_r}{dr} + \frac{4\pi r P_r^2}{1-2m/r} +
\frac{P_r m}{r^2 \left(1-2m/r\right)} +
\frac{4\pi r \rho P_r}{1-2m/r} + \nonumber \\
 &  & + \frac{\rho m}{r^2 \left(1-2m/r\right)} -
\frac{2\left(P_\bot - P_r\right)}{r} \nonumber \\
& = & P'_r + \frac{\nu'}{2} \left(\rho + P_r\right) -
\frac{2}{r} \left(P_\bot - P_r\right)
\label{Rfr}
\end{eqnarray}

\noindent
The physical meaning of $R$ is clearly inferred from (\ref{Rfr}).
It represents the total force (gravitational + pressure gradient +
anisotropic term) acting on a given fluid element. Obviously,
$R>0/R<0$ means that the total force is directed $inward/outward$ of
the sphere.

\noindent
Let us now turn back to thermal conduction equation (\ref{Catrel}).
Evaluating it
immediately after perturbation, we obtain for both initial
configurations (static and slowly evolving) (see \cite{Heetal97} for details)

\begin{equation}
\tau \dot Q e^{\lambda/2} = - \kappa T \dot\omega
\label{Cat1}
\end{equation}

\noindent
Finally, combining (\ref{pfR}) and (\ref{Cat1}), one obtains

\begin{equation}
\dot\omega = - \frac{e^{(\nu-\lambda)/2} R}{\left(\rho+P_r\right)}
\times
\frac{1}{\left(1 - \frac{\kappa T}{\tau \left(\rho+P_r\right)}\right)}
\label{exmin}
\end{equation}

\noindent
or, defining the parameter $\alpha$ by

\begin{equation}
\alpha \equiv \frac{\kappa T}{\tau \left(\rho + P_r\right)}
\label{alfa}
\end{equation}

\begin{equation}
- e^{(\nu-\lambda)/2} R =
\left(\rho + P_r\right) \dot \omega \left(1 - \alpha\right)
\label{Ralfa}
\end{equation}

\noindent
This last expression
has the obvious ``Newtonian'' form

\centerline{Force $=$ mass $\times$ acceleration}

\noindent
since, as it is well known, $\left(\rho + P_r\right)$ represents
the inertial mass density and by ``acceleration'' we mean the time derivative
of $\omega$ and not $(a_\mu a^\mu)^{1/2}$. If $\alpha<1$, then an
$outward/inward$ acceleration ($\dot \omega>0/\dot \omega<0$) is
associated with an $outwardly/inwardly$ ($R<0/R>0$) directed total
force (as one expects!).

\noindent
However, if $\alpha=1$, we obtain that
$\dot \omega \not = 0$ even though $R=0$. Still worse, if
$\alpha>1$, then an $outward/inward$ acceleration is associated with an
$inwardly/outwardly$ directed total force!.

\noindent
As mentioned before, the critical point may be restricted by causality
conditions,
particularly in the pure bulk or shear viscosity case \cite{HeMa972},
however this
is not so in the general case
\cite{HeMa972}.

\noindent
Independently of this fact, it is clear from (\ref{Ralfa}),
that the ``effective'' inertial mass term decreases  as $\alpha$
increases.
\noindent
In the next section we shall obtain an expression for the active
gravitational mass
explicitly containing $\alpha$.

\section{The Tolman mass}

\noindent
The Tolman mass for a spherically symmetric distribution
of matter is given by (eq.(24) in \cite{To})

\begin{eqnarray}
m_T = & &  4 \pi \int^{r_\Sigma}_{0}{r^2 e^{(\nu+\lambda)/2}
\left(T^0_0 - T^1_1 - 2 T^2_2\right) dr}\nonumber \\
& + & \frac{1}{2} \int^{r_\Sigma}_{0}{r^2 e^{(\nu+\lambda)/2}
\frac{\partial}{\partial t}
\left(\frac{\partial L}{\partial \left[\partial
\left(g^{\alpha \beta} \sqrt{-g}\right) / \partial t\right]}\right)
g^{\alpha \beta}dr}
\label{Tol}
\end{eqnarray}

\noindent
where $L$ denotes the usual gravitational lagrangian density
(eq.(10) in \cite{To}). Although Tolman's formula was introduced
as a measure of the total energy of the system, with no commitment
to its localization, we shall define the mass within a sphere of
radius $r$, inside $\Sigma$, as

\begin{eqnarray}
m_T = & &  4 \pi \int^{r}_{0}{r^2 e^{(\nu+\lambda)/2}
\left(T^0_0 - T^1_1 - 2 T^2_2\right) dr}\nonumber \\
& + & \frac{1}{2} \int^{r}_{0}{r^2 e^{(\nu+\lambda)/2}
\frac{\partial}{\partial t}
\left(\frac{\partial L}{\partial \left[\partial
\left(g^{\alpha \beta} \sqrt{-g}\right) / \partial t\right]}\right)
g^{\alpha \beta}dr}
\label{Tolin}
\end{eqnarray}

\noindent
This (heuristic) extension of the global concept of energy to a local level
\cite{Coo} is suggested by the conspicuous role played by
$m_T$ as the ``effective gravitational mass'', which will be
exhibited below.

\noindent
On the other hand, even though Tolman's definition is not
without its problems \cite{Coo,Deu}, we shall see that $m_T$,
as defined by (\ref{Tolin}), is a good measure of the
active gravitational mass, at least for the system under
consideration.

\noindent
After some simple but tedious calculations, it can be shown that (\ref{Tolin})
may be written as (see \cite{DPHeHPSa} for details).

\begin{equation}
m_T = e^{(\nu + \lambda)/2}
\left[m(r,t) - 4 \pi r^3 T^1_1\right]
\label{I+II}
\end{equation}

\noindent
where the mass function $m(r,t)$ is defined by \cite{MiSh,CaMc}

\begin{equation}
m(r,t) = \frac{1}{2} r R^3_{232}
\label{mR}
\end{equation}

\noindent
and the Riemann component for metric (\ref{metric}) is
given by

\begin{equation}
R^3_{232} = 1 - e^{-\lambda}
\label{R32}
\end{equation}

\noindent
Using field equations, (\ref{mR}) may be written in the most familiar form

\begin{equation}
m(r,t) = 4 \pi \int^r_0{r^2 T^0_0 dr}
\label{mT00}
\end{equation}

\noindent
or alternativately \cite{DPHeHPSa}

\begin{equation}
m(r,t) = \frac{4 \pi}{3} r^3 \left(T^0_0 + T^1_1 - T^2_2\right) +
\frac{r}{2} C^3_{232}
\label{mTTT}
\end{equation}

\noindent
where $C^3_{232}$ denotes the corresponding component of the Weyl tensor.

\noindent
It is worth noticing that this is, formally, the same
expression for $m_T$ in terms of $m$ and $T^1_1$, that
appears in the static (or quasi-static) case
(eq.(25) in \cite{HeSa95}).

\noindent
Replacing $T^1_1$ by (\ref{feq11}), and $m$ by (\ref{mR}) and (\ref{R32}),
one may also obtain

\begin{equation}
m_T = e^{(\nu - \lambda)/2} \, \nu' \, \frac{r^2}{2}
\label{mT}
\end{equation}

\noindent
This last equation brings out the physical meaning of $m_T$ as the
active gravitational mass. Indeed, it can be easily shown \cite{Gro}
that the gravitational acceleration ($a$) of a test particle,
instantaneously at rest in a static gravitational field, as measured
with standard rods and coordinate clock is given by

\begin{equation}
a = - \frac{e^{(\nu - \lambda)/2} \, \nu'}{2} = - \frac{m_T}{r^2}
\label{a}
\end{equation}

\noindent
A similar conclusion may be obtained by inspection of eq.(\ref{Prp})
(valid only in the static or quasi-static case) \cite{Lig}.
In fact, the first term on the right side of this equation
(the ``gravitational force'' term) is a product of the ``passive''
gravitational mass density $(\rho + P_r)$ and a term proportional
to $m_T/r^2$.

\noindent
We shall now consider another expression for $m_T$,
which appears to be more suitable for the treatment of
the problem under consideration. This latter expression
will be evaluated immediately after the system departs
from equilibrium.
Therefore the physical meaning of $m_T$ as the active
gravitational mass obtained for the static (and quasi-static)
case, may be safely extrapolated to the non-static case within
the time scale mentioned above.

\noindent
The required expression for the Tolman mass will be obtained as follows
(see \cite{DPHeHPSa} for details).
Taking the $r$-derivative of (\ref{mT}) and using (\ref{mTTT}) and (\ref{I+II})
we obtain the following differential equation for $m_T$

\begin{eqnarray}
r m'_T - 3 m_T & = & e^{(\nu + \lambda)/2} \left[
4 \pi r^3 \left(T^1_1 - T^2_2\right) - 3 W_{(s)}\right] \nonumber \\
& + &
\frac{e^{(\lambda - \nu)/2} r^3}{4} \left(\ddot\lambda +
\frac{\dot\lambda^2}{2} - \frac{\dot\lambda \dot\nu}{2}\right)
\label{pre}
\end{eqnarray}

\noindent
where $W_{(s)}$ is given by

\begin{equation}
W_{(s)} =
\frac{r^3 e^{-\lambda}}{6}
\left( \frac{e^\lambda}{r^2} - \frac{1}{r^2} +
\frac{\nu' \lambda'}{4} - \frac{\nu'^2}{4} -
\frac{\nu''}{2} - \frac{\lambda'}{2r} + \frac{\nu'}{2r} \right)
\label{Ws}
\end{equation}

\noindent
Equation (\ref{pre}) can be formally integrated to obtain

\begin{eqnarray}
m_T & = & (m_T)_\Sigma \left(\frac{r}{r_\Sigma}\right)^3 \nonumber \\
& - & r^3 \int^{r_\Sigma}_r{e^{(\nu+\lambda)/2} \left[\frac{8 \pi}{r}
\left(T^1_1 - T^2_2\right)
+ \frac{1}{r^4} \int^r_0{4 \pi \tilde{r}^3 (T^0_0)' d\tilde{r}}
 \right] dr} \nonumber \\
& - & r^3 \int^{r_\Sigma}_r{
\frac{e^{(\lambda-\nu)/2}}{2r}\left(\ddot\lambda + \frac{\dot\lambda^2}{2}
- \frac{\dot\lambda \dot\nu}{2}\right)
  dr}
\label{emte}
\end{eqnarray}

\noindent
In the static (or quasi-static) case
($\ddot\lambda = \dot\lambda^2 = \dot\lambda \dot\nu = 0$)
the expression above is identical to eq.(32) in \cite{HeSa95}.

\noindent
We shall now proceed to evaluate (\ref{emte}) immediately after
perturbation.
Using (\ref{T00})--(\ref{T01}) and (\ref{omyQ0})--(\ref{pomQ2}),
we see that up to first order we get immediately after perturbation
(for both initial conditions)

\begin{eqnarray}
& & T^0_0 = \rho \, \qquad \, T^1_1 = - P_r \, \qquad \,
T^2_2 = - P_\bot \nonumber \\
& & \dot\lambda^2 = \dot\nu \dot\lambda = 0 \nonumber \\
& & \ddot\lambda = - 8 \pi r e^{(\nu + \lambda)/2}
\left[\left(\rho + P_r\right) \dot\omega + \dot Q e^{\lambda/2}\right]
\label{T=}
\end{eqnarray}

\noindent
Replacing (\ref{T=}) into (\ref{emte}) and using (\ref{Cat1}) and
(\ref{Ralfa}),
we obtain finally

\begin{eqnarray}
m_T & = & (m_T)_\Sigma \left(\frac{r}{r_\Sigma}\right)^3 \nonumber \\
& + & 4 \pi r^3 \int^{r_\Sigma}_r{e^{(\nu + \lambda)/2}
\left[\frac{2}{r} \left(P_r - P_\bot\right) -
\frac{1}{r^4} \int^r_0{\tilde{r}^3 \rho' d\tilde{r}}\right] dr} \nonumber \\
& + & 4 \pi r^3 \int^{r_\Sigma}_r
{e^{\lambda} \left(\rho + P_r\right) \dot\omega \left(1 - \alpha\right) dr}
\label{mfi}
\end{eqnarray}

\noindent
where the general expression for $\left(m_T\right)_\Sigma$ can be obtained from
(\ref{I+II}), (\ref{T11}), (\ref{defq}) and (\ref{PQ})

\begin{equation}
\left(m_T\right)_\Sigma = m_\Sigma + \frac{4 \pi r^3_\Sigma \hat{q}_\Sigma
\left(1 + 2 \omega_\Sigma\right)}{1 - \omega^2_\Sigma}
+ 4 \pi r^3_\Sigma \left(\frac{\rho \omega^2}{1 - \omega^2}\right)_\Sigma
\label{mtsig}
\end{equation}

\noindent
which, after perturbation reduces to

\begin{equation}
\left(m_T\right)_\Sigma = m_\Sigma + 4 \pi r^3_\Sigma \hat{q}_\Sigma
\label{mtsaf}
\end{equation}

\section{Discussion}

\noindent
Let us now consider a sphere of radius $r$ within $\Sigma$. Immediately
after perturbation
the Tolman mass of this internal core is given by (\ref{mfi}).
The relevance of the two terms in the first integral has already been
discussed \cite{DPHeHPSa}
and therefore they shall not be considered here.

\noindent
Instead, we shall focus on the last term in (\ref{mfi}).
If the system starts to collapse $(\dot\omega<0)$ this last term tends to
decrease
the value of the Tolman mass, leading thereby to a weaker collapse.
Inversely, if the system starts to expand $(\dot\omega>0)$, the last term
in (\ref{mfi})
contributes positively to the Tolman mass of the core, leading to a weaker
expansion.
Thus, in both cases this last term tends to stabilize the system.
This is so as long as $\alpha<1$.

\noindent
If $\alpha>1$ the inverse picture follows.
In this case for initially collapsing (expanding)
configurations the last term in (\ref{mfi}) becomes positive (negative)
leading to stronger
collapse (expansion).

\noindent
In general the system becomes more and more unstable
 as $\alpha$ grows.

\noindent
It should be noticed that in the comments above we have assumed $\alpha$ to
be constant
througout the fluid distribution.
This of course is a rather crude approximation as it is evident from
(\ref{alfa}).
Therefore, a wide variety of scenarios may be considered from different radial
dependence of that parameter.

\noindent
Finally, observe that in the dissipationless case ($\alpha = 0$) ,
an inflationary equation of state $(\rho = - P_r)$ is equivalent to
the critical point ($\alpha = 1$) in the heat conducting situation.
In both cases the stabilizer term in (\ref{mfi}) vanishes.

\noindent
In the heat conducting case $\alpha \not = 0 $,
an inflationary equation of state leads to
an effective inertial mass density equal to $- \kappa T/\tau$, as follows
from (\ref{pfR}) and (\ref{Cat1}).
In this case, according to (\ref{Cat1}), (\ref{emte}) and (\ref{T=})
the last integral in (\ref{mfi}) should be replaced by

\begin{equation}
- 4 \pi r^3 \int^{r_\Sigma}_{r}{e^\lambda \frac{\kappa T}{\tau} \dot\omega dr}
\end{equation}

\noindent
This last integral contributes negatively to Tolman mass in the case
$\dot\omega > 0$,
yielding stronger expansions.

\noindent
In other words, in what concerns eq.(\ref{mfi}), an equation of state of
the above mentioned
form $(\rho = - P_r)$ is equivalent (in the dissipative case) to a
situation with $\alpha > 1$.

\noindent
Of course one might ask if a real physical system may reach (or even go
beyond) the critical point.
The answer to this question seems to be affirmative as suggested by the
example provided in \cite{HeMa972}.
Indeed it is shown in that reference that a mixture of matter and neutrinos
with typical values of temperature
and energy density, corresponding to the moment of birth of a neutron star
in a supernova explosion may lead to
values of $\alpha$ equal to or even greater than $1$.

\noindent
However, it is not our purpose here to discuss about the plaussibility to
reach the critical point
but rather to bring out the physical meaning of $\alpha$ and the critical point.

\noindent
Finally it is worth noticing that evaluating the mass function from
(\ref{I+II}) and using (\ref{mfi}),
we obtain similar conclusions about the relation between $\alpha$ and the
mass function as those obtained
for the Tolman mass.

\noindent
However, unlike the Tolman expression, the mass function can not be
interpreted (for a part of the configuration)
as the active gravitational mass and therefore the stability/instability
criteria look less convincing
when using $m(r,t)$ instead of $m_T(r,t)$.

%\newpage

\end{document}